\documentclass[12pt]{article}%
\usepackage{amsmath}
\usepackage{graphicx}
\usepackage{wrapfig}
\usepackage{lscape}
\usepackage{rotating}
\usepackage{epstopdf}
\usepackage{amsfonts}
\usepackage{amssymb}
\usepackage{color}%
\usepackage{tikz}
\usetikzlibrary{shapes,arrows}

\usepackage[colorinlistoftodos]{todonotes}

\begin{document}
\title{Prediction and forecasting models based on patient's history
and biomarkers with application to Scleroderma disease}
\author{
Haiyan Liu$^1$, Francesco Del Galdo$^{2,3}$ and Jeanine Houwing-Duistermaat$^{1,4}$
\\$^1$Department of Statistics, University of Leeds
\\$^2$Institute of Rheuamtic and Musculoskeletal Medicine, University of Leeds
\\$^3$NIHR Biomedical Research Centre, Leeds Teaching Hospital Trusts
\\$^4$Department of Biostatistics and Research Support, \\ Julius Center, UMC Utrecht}
\maketitle

\begin{abstract}
This paper aims at predicting lung function values based on patients historical lung function values and serum biomarkers in Scleroderma patients. The progression of disease is  measured by three lung function indexes (FVC, TLC, DLCO). Values of four biomarkers (TIMP1, P3NP, HA, NT-proBNP) are available. The data are sparse (6 months intervals) and irregular (many visits are missed). We consider two modeling approaches to achieve our goal, namely, the mixed effects model which is the standard approach in epidemiological studies and the functional principal component analysis model which is typically used for dense temporal datasets.
We find that functional data methodology was able to recover the trajectories of three lung function indexes and to predict the future values very well.
\end{abstract}

Keywords: functional principal component analysis, mixed effects model,
prediction, Scleroderma, sparse and irregular functional data.

\section{Introduction}
With the introduction of the Electronic Health Records (EHR), it is now possible to use 
retrospective information on outcomes, biomakers and classical risk factors to build 
a prediction model for a specific outcome. 
These models can be used for decision making with regards to planning of hospital visits, 
expensive measurements and treatment. 
Especially for diseases with heterogeneous trajectories, 
i.e.\ where the progress and type of the disease differ across patients, 
patient specific prediction is relevant. 
Unfortunately, building models from EHR data is often complex. 
The visits of patients are typically irregular and sparse and the number of patients 
with regular observations might be small.  
The heterogeneity in disease and underlying risks needs to be taken into account. 
Finally the data are typically noisy.

In epidemiology linear mixed models are often used to model the relationship between biomarkers and
a repeatedly measured outcome. 
Here correlation between measurements for the same person is modeled via random effects. 
For mixed effects model theory see 
Verbeke and Molenberghs (2009),  Hedeker and Gibbons (2006), 
Wu and Zhang (2006), Wu and Tian (2018) and the references therein.
Forecasting of future disease trajectories based on the current values of the risk factors, 
covariates and outcomes might be performed with these models, 
see Skrondal and Rabe-Hesketh (2009). 

Alternative approaches are in the framework of functional data analysis. 
The advantages of these latter methods are that they are very flexible with regard to the functional 
model for the outcome over time and they can deal with irregular and sparse data,
see e.g. Beran and Liu (2014, 2016), Peng and Paul (2009),
Yao et al. (2005), James (2010), James et al. (2000). 
For general methods and theory on functional data analysis, one may refer to the monographs 
such as Ramsay and Silverman (2005), Horvath and Kokoszka (2012), Kokoszka and Reimherr (2017) 
and the references therein. 
In the FDA framework estimation of future disease trajectories is straightforward, in contrast to predictions based on a mixed model, which requires integration over the posterior multivariate distribution of the random effects. On the other hand, a mixed model provides a straightforward prediction of the outcome by just filling in the covariate values and zeros for the random effects. 
This paper is motivated by an ongoing study on predicting the disease trajectory of Scleroderma patients. Here we discuss methods for prediction of expensive measurements based on historical and biomarker data. We will consider linear mixed models and functional data analysis for this purpose.



We will build prediction models for Scleroderma disease. 
Scleroderma disease is a rare, clinically heterogeneous multisystem disorder
which might affect the patients' physical and psychological
functioning and impairs their ability to
participate in work and social activities (see Jaeger et al. 2017, 
Gabrielli et al. 2009, and Muangchan et al. 2013). 
The disease is heterogeneous since only 30\% of the patients progress to a severe disease status. 
Further the patients can progress to various outcomes. 
Currently, the most common outcomes of disease progression include
modified rodnan skin score (mRSS) and three lung functions 
(FVC, TLC, DLCO) which are age and gender specific predicted percentages.
In this paper, we will not consider mRSS since it is quite noisy and depends on the observer (see Khanna et al. 2017 and
Clements et al. 1993). 
Since the lung functions are expensive measurements, good predictions of these functions based on 
historical or biomarker data will be valuable. 
Further, we have data on four candidate biomarkers, namely TIMP1, P3NP, HA, 
and NT-proBNP.
In the rest of the paper, we denote TIMP1 by TIMP and NT-proBNP by NT.

Data for this study come from outpatient EHRs of patients attending Scleroderma clinics every six months. 
Scleroderma patients visit the hospital every six months. In our dataset the number of patients with regular visits is very limited. On the other hand we have a few patients with several measurements over a longer time period. To assess prediction accuracy based on current values of the biomarkers we will use linear mixed models for a subset of patients with at least two observations
at 0, 6, 12 months. 
Since the outcomes change continually over time, they are functional data. 
We propose therefore to use a functional principal component analysis (FPCA) method to predict the 
future values of outcomes.
Here we use the subset of patients who have at least two observations in the first 60 months.
Restricted maximum likelihood framework is used to estimate the eigenelements of 
underlying covariance operator (see Peng and Paul 2009) and the scores are estimated 
by the principal components analysis through conditional expectation (PACE) method 
(see Yao et al. 2005).
Then the trajectories are recovered by using the truncated Karhunen-Lo\`eve decomposition based 
on the estimated eigenelements and scores.

In section 2 we describe the dataset in more detail. 
In Section 3 we will introduce the methods. 
Section 4 presents the results. 
Finally section 5 is a conclusion and future research lines are provided.

\section{Dataset}
\begin{figure}
\tikzstyle{block} = [rectangle, draw, 
text width=18em, text centered, rounded corners, node distance=1.6cm, minimum height=2em]
\tikzstyle{cloud} = [rectangle, draw, node distance=7.0cm, 
text width=12em,  text centered, rounded corners, minimum height=2em]
\tikzstyle{line} = [draw, -latex']

\begin{tikzpicture}[node distance = 2cm, auto]
\node [block] (Scl) 
{Scleroderma study,\\ \# patients=255, \# obs=573};
\node [block, below of=Scl] (NA) 
{ remove NA observations \\ \# patients=208, \# obs=421};
\node [block, below of=NA] (TIMPlager500less50) 
{remove patients with TIMP(${t_0}$) $>500$ and TIMP(${t_0}$) $<50$ , 
\\ \# patients=206, \# obs=418};
\node [block, below of=TIMPlager500less50] (more60) 
{remove observations of $t>60$, \\ \# patients=206, \# obs=416};
\node [block, below of=more60] (oneObs) 
{remove patients with one observation, \\ \# patients=123, \# obs=333};
\node [block, below of=oneObs] (P3NPlager25) 
{remove observations with P3NP $>25$, 
\\ \# patients=123, \# obs=332};
\node [block, below of=P3NPlager25] (FVClager170less25) 
{remove observations with FVC $>170$ and FVC $<50$, 
\\ \# patients=122, \# obs=329};
\node [block, below of=FVClager170less25] (TLCless25) 
{remove observations with TLC $<25$, 
\\ \# patients=122, \# obs=328};
\node [block, below of=TLCless25] (DLCOlarger120) 
{remove observations with DLCO $<120$, 
\\ \# patients=121, \# obs=326};
\node [block, below of=DLCOlarger120] (NTless1000) 
{remove observations with NT $<1000$, 
\\ \# patients=117, \# obs=316};
\node [cloud, right of=NTless1000] (badNT) 
{dataset for FDA modelling};
\node [block, below of=NTless1000] (LMMObs) 
{patients with at least 2 observations at 0, 6, 12 months, \\ \# patients=73, \# obs=159};
\node [cloud, right of=LMMObs] (pred) {dataset for LMM};
\path [line] (Scl) -- (NA);
\path [line] (NA) -- (TIMPlager500less50);
\path [line] (TIMPlager500less50) -- (more60);
\path [line] (more60) -- (oneObs);
\path [line] (oneObs) -- (P3NPlager25);
\path [line] (P3NPlager25) -- (FVClager170less25);
\path [line] (FVClager170less25) -- (TLCless25);
\path [line] (TLCless25) -- (DLCOlarger120);
\path [line] (DLCOlarger120) -- (NTless1000);
\path [line, dashed] (NTless1000) -- (badNT);
\path [line] (NTless1000) -- (LMMObs);
\path [line, dashed] (LMMObs) -- (pred);
\end{tikzpicture}
\caption{Patient chart flow.}
\label{figure0PatientFlow}
\end{figure}
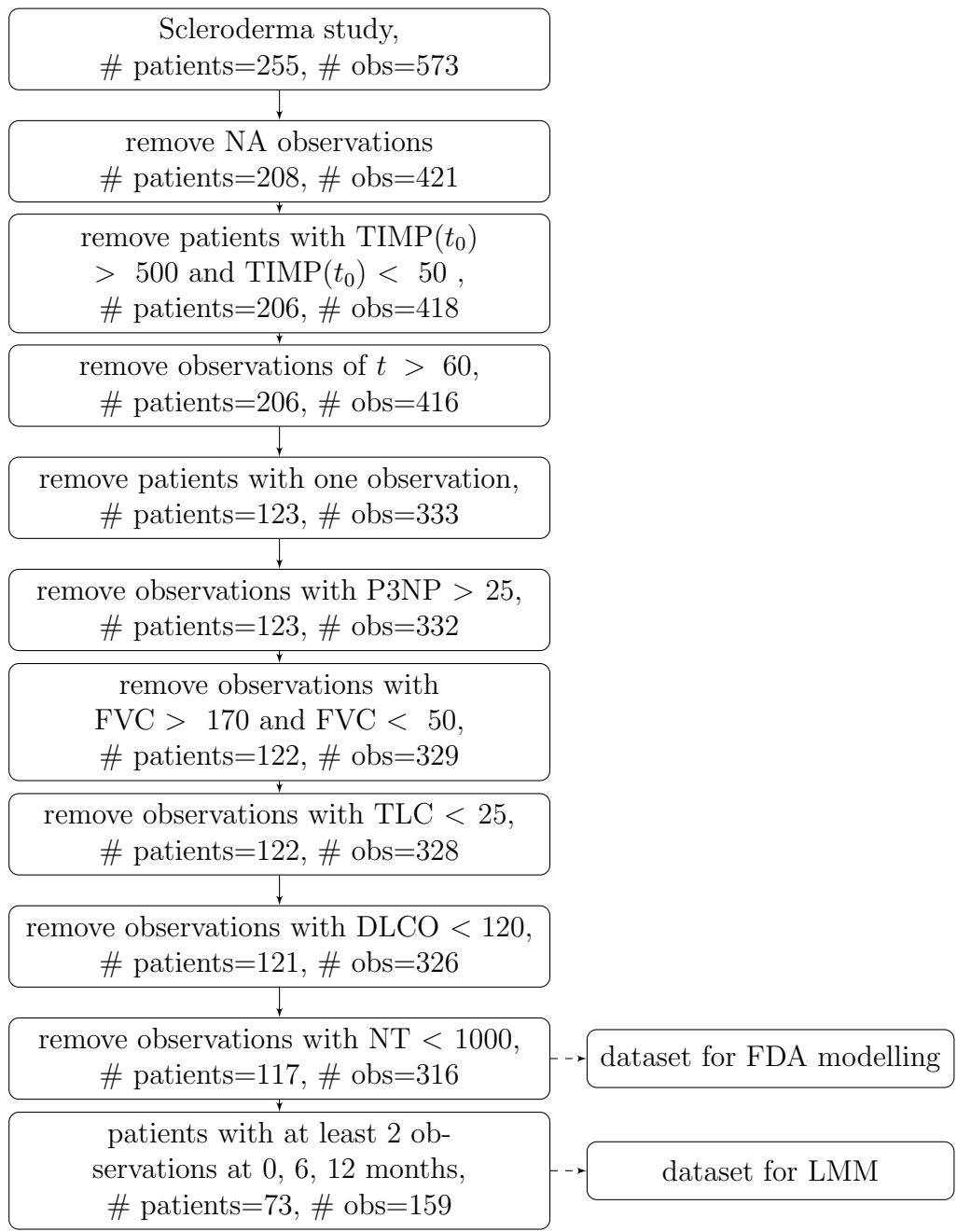

In total, 255 Scleroderma patients with information from hospital visits from 
1995 to 2015 have been retrospectively recruited to our study,
the majority of the observations are from after 2010).
Data were collected according to ethically approved protocol
for observational study  HRA number 15/NE/0211.
Typically, the Scleroderma patients visit hospital every six months.
However, many patients missed their appointments or data were not recorded.
Therefore, our dataset has only 573 observations.
We have three lung function indexes FVC, TLC, DLCO with values larger than 0 and small values
for severe disease.
Four biomarkers are measured, namely TIMP, P3NP, HA and NT.
After removing the observations with missing values, our dataset comprises 
421 measurements from 208 patients. 
Further data cleaning results in 316 measurements from 117 patients see Figure \ref{figure0PatientFlow}
for details.

First for mixed effects models, we select the patients who have at least two observations at 
0, 6, 12 months and the number of patients is 73.
Second for the functional data analysis model, we use all the 117 patients for modelling.
For forecasting, we need patients at least three observations in the first 60 months and
the number of patients is 60.

\begin{figure}
  \includegraphics[angle=-90, width=0.88\textwidth]{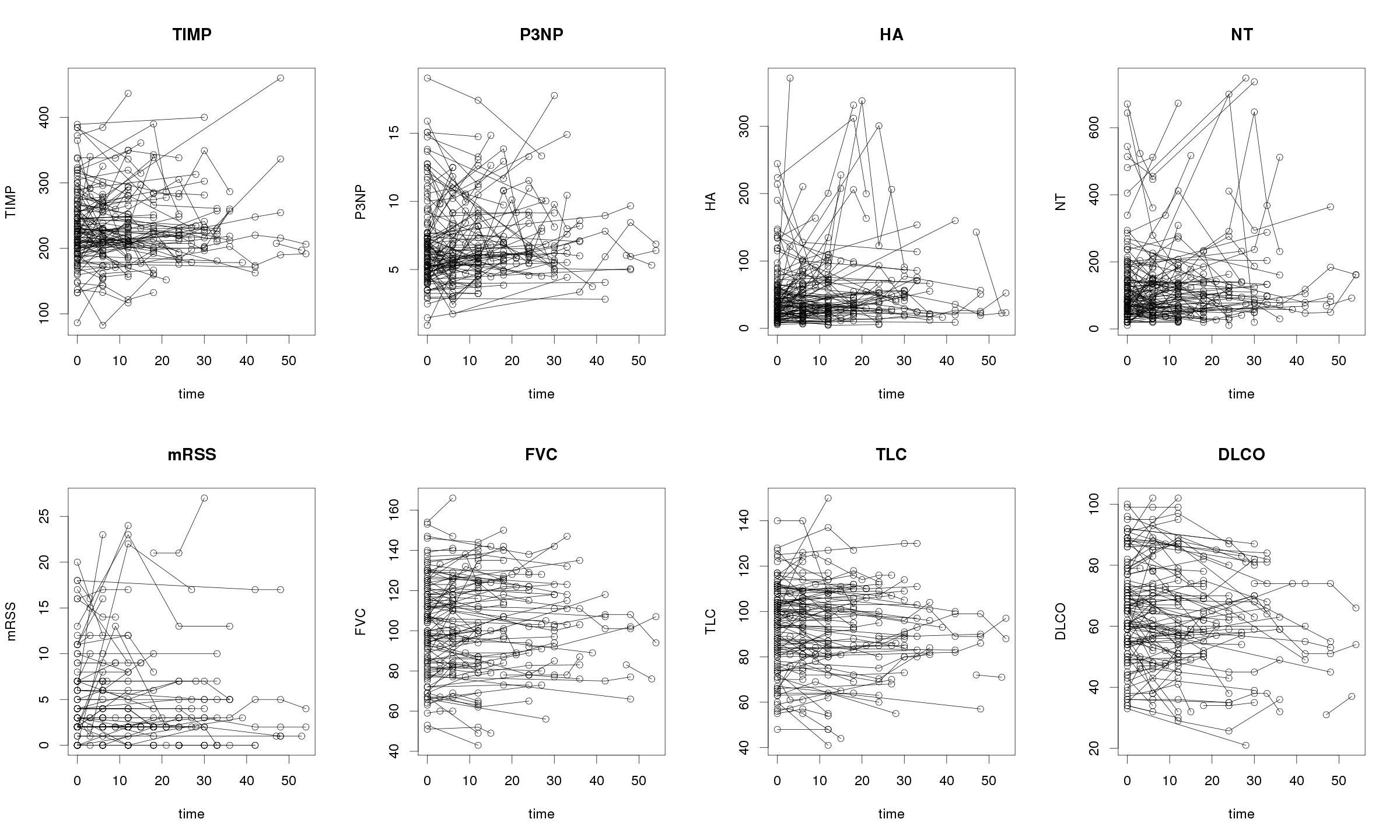}
  \caption{Clean data.}
  \label{FigureCleanData}
\end{figure}

The profiles of these patients with regard to the biomarkers and outcomes are depicted in Figure
\ref{FigureCleanData}.
Figure \ref{FigureCleanData} also shows that the observations are sparse, 
irregular and unbalanced, with more observations in the first two years and less in the last two years.
From these patients, only 60 patients (with 202 measurements) have data for at least three visits. For time 0, i.e. the first visit time, the mean and covariance of biomarker 
and outcomes are shown in Table \ref{Table-Description}.
The correlations between FVC, TLC and DLCO are quite large, with the largest value of 0.80 for the correlation between FVC and TLC. The correlations between the biomarkers are smaller and range from 0.15 to 0.44.

\begin{table}[ptbh]
\centering  {
\begin{tabular}
[c]{|c|c|c|c|c|c|c|c|c|}\hline
& TIMP & P3NP & HA & NT & FVC & TLC & DLCO \\\hline
 \multicolumn{8}{|c|}{mean}\\\hline
 & 234 & 7  & 49 & 130 & 102 & 93 & 65 \\\hline
 \multicolumn{8}{|c|}{variance}\\\hline
 & 3188 & 11 & 2098 & 14875  & 540 & 329 & 267 \\\hline
 \multicolumn{8}{|c|}{correlation}\\\hline
TIMP & 1.00 & 0.35 & 0.30 & 0.21  & -0.21 & -0.28 & -0.19  \\
P3NP &      & 1.00 & 0.44 & 0.15 & -0.32 & -0.35 & -0.36 \\
HA   &      &      & 1.00 & 0.15  & 0.05 & -0.08 & -0.04 \\
NT   &  &  &  & 1.00  & -0.08 & -0.21 & -0.21 \\
FVC  &  &  &  & &   1.00 & 0.80 & 0.47 \\
TLC  &  &  &  & &   & 1.00 & 0.51 \\
DLCO &  &  &  & &  &  & 1.00 \\
\hline
\end{tabular}
}
\caption{At time 0 (112 observations), mean and variance of biomarkers and outcomes, 
correlations between biomarkers and outcomes.
}
\label{Table-Description}
\end{table}

\section{Methods}
Our aim is to predict and forecast the lung functions  by using either biomarkers or past lung function values. 
The classical way to analyse sparse longitudinal data is the linear mixed effects model. 
The grid of visits should be regular although missing outcomes for some visits are allowed. 
Alternatively sparse and irregular longitudinal data are functional data,
and forecasting based on historical outcomes can be performed by using functional principal component analyses. 
For our dataset we will consider both methods.



\subsection{Linear mixed effects models}
To model the relationship between biomarkers and lung function indexes, linear mixed
effects models are fitted to the data from patients with observations at time 0 ($n=73$) 
and at least one observation at 6 months ($n=42$) or at 12 months ($n=44$).


For individual $i$ ($i=1,...,73$) and measurement $j$ ($j=1, ..., N_i $ with $N_i\in \{2,\ 3\}$ 
the number of observations for individual $i$), let $y_{ij}$ be the observation at time 
$t_{ij}$, $x_{ij}$ be the logarithm of the value of the biomarker 
for individual $i$ at time $t_{ij}$. 
Then the following linear mixed effects model is considered
\begin{equation}
\label{Equation-Mixed}
y_{ij}=\beta_{0}+\beta_1 t_{ij}+\beta_{2} x_{ij}+\gamma_{0i}+\gamma_{1i} t_{ij}+e_{ij}
\end{equation}
where $t_{ij}=0, 6$ or 12 months, $\beta_{0j}+\beta_1 t_{ij}$ is the time specific intercept 
and $\beta_{2j}$ represents the effect of the biomarker at time point $t_{ij}$. 
The the person specific random intercept $\gamma_{0i}$ and slope $\gamma_{1i}$ model follow the following distribution
$$(\gamma_{0i}, \gamma_{1i})^T\sim N(0, \Sigma)$$
with
$$\Sigma=
\left( {\begin{array}{cc}
   \sigma^2_0 & \sigma_{01} \\
   \sigma_{01} & \sigma^2_1 \\
  \end{array} } \right).
  $$
Finally the term $e_{ij}$ is the random residual.
Note that since it is expected that heterogeneity across the patients with regard to disease progress exist we included a random intercept and slope.

Based on the linear mixed model, a prediction for a new lung function value can be computed based 
on current biomarkers by plugging in the mean of the random effects (i.e. 0) 
in (\ref{Equation-Mixed}) (see Skrondal and Rabe-Hesketh 2009). 
To evaluate the prediction accuracy of this model, the cross validated mean squared error (MSE) is computed. 
To this end we base the prediction of the vector of outcomes $y_{i}$ on the vector of biomarkers $x_{i}$ 
and a model fitted on the data set obtained by leaving out the data of indivivual $i$. 
Specifically
\begin{equation*}
\hat y_{ij}=\hat\beta^{(-i)}_0+\hat\beta^{(-i)}_1 x_{ij}+\hat\beta^{(-i)}_2 t_{ij}
\end{equation*}
where $\hat\beta^{(-i)}_0, \hat\beta^{(-i)}_1, \hat\beta^{(-i)}_2$ are the corresponding estimates of 
$\beta_0, \beta_1, \beta_2$ of model (\ref{Equation-Mixed}) on the data set obtained by leaving out 
the observations of individual $i$.
Now the leave one out cross validated average MSE is defined as:
\begin{equation*}
MSE = \frac{1}{\sum_{i=1}^n N_i}\sum_{i=1}^n\sum_{j=1}^{N_i} ( y_{ij}- \hat y_{ij})^2.
\end{equation*}

In principle linear mixed models can also be used to forecast a new measurement in an existing cluster. 
However since in our dataset we have only a few patients ($n=13$) who have 
three observations at 0, 6, 12 months, 
the performance of these predictions cannot be evaluated.
An alternative is functional principal component analysis.


\subsection{Functional principal component analysis}
Let $X_i(t)$ be the trajectory of the $i$th patient, where $i\in\{1,...,n\}, \ t\in[0,1]$.
Here the time scale is transformed from 0 to 60 months to $[0,1]$.
We assume $X_i(t)$ are independent realizations of a squared integrable random function $X(t)$ 
on $[0, 1]$, i.e. $X(t)\in L^2[0, 1]$.
We denote $\mu(t)=E[X(t)]$ the mean function of $X(t)$ and $C(s,t)=E[(X(s)-\mu(s))(X(t)-\mu(t))]$
the covariance function of  $X(t)$, then by Mercer's theorem
\begin{align*}
C(s,t)=\sum_{l=1}^\infty\lambda_l\phi_l(s)\phi_l(t)
\end{align*}
where $\lambda_1\geq\lambda_2\geq...\geq0$ with $\sum\lambda_l<\infty$ are the eigenvalues of $C$.
$\phi_l(t)$ is the $l$th eigenfunction corresponding to $\lambda_l$ and $\{\phi_l\}$ forms
a complete orthonormal basis for $L^2[0, 1]$. 
Then $X_i(t)$ has the following Karhunen-Lo\`eve decomposition
\begin{align*}
X_i(t)=\mu(t)+\sum_{l=1}^\infty\xi_{il}\phi_l(t)
\end{align*}
where $\xi_{il}=\int(X_i(t)-\mu(t))\phi_l(t)dt$ is the associated functional principal score with
$var(\xi_{il})=\lambda_l$ and $\phi_l$ is also called the $l$th functional principal component.
In practice, $X_i(t)$ is usually well approximated by the first $L$ 
leading functional principal components and scores
\begin{align*}
\tilde X_i(t)=\mu(t)+\sum_{l=1}^L\xi_{il}\phi_l(t)
\end{align*}

For patient $i$ the trajectory $X_i(t)$ is observed at a sparse grid $t_{i1}, ...t_{iN_i}$. 
For example, $X_i(t_{i1}), ..,X_i(t_{i_{N_i}})$ are the observed 
discrete DLCO values for patient $i$ at time $t_{i1}...,t_{i_{N_i}}$ (after rescaling).
The number and spacing of the observed time points vary per patient, i.e.
for $i\ne j$, $N_i$ and $N_j$ might be different and $t_{i2}-t_{i1}$
might be different from $t_{j2}-t_{j1}$. 

In order to recover the individual-specific curves $X_i(t)$, one has to 
estimate the mean function $\mu(t)$, FPC scores $\xi_{il}$, FPCs $\phi(t)$,
and determine the number of FPCs to be included $L$.
The mean function $\mu(t)$ is estimated by a local linear smoother
and it is subtracted from the observations before estimating the other model parameters.

Since the eigenfunctions corresponding to the largest several eigenvalues can be well approximated
by using less complex basis, cubic B-spline basis with equally spaced knots is used to expand
$\phi_l$.
Under the Gaussianity of the process, i.e. $\xi_{il}\overset{iid}\sim N(0, \lambda_l)$
and the errors $\epsilon$ are normal, 
the coefficients of expansion of $\phi_l$, eigenvalues $\lambda_l$ and other nuisance parameters
can be estimated through the restricted maximum likelihood procedure proposed 
by Peng and Paul (2009). 
The number of nonzero eigenvalues and the number of basis for representing the eigenfunctions are
selected by leave-one-curve-out cross-validation criteria.

Finally, for each patient, we need to estimate the FPC scores.
The traditional way to estimate the FPC scores $\xi_{il}=\langle X_i-\mu, \phi_l\rangle$ 
is through numerical integration if the sampling is dense, see e.g. Beran and Liu (2016).
However, for sparse functional data, numerical integration estimator will not be a 
reasonable estimate of $\xi_{il}$. 
Under the Gaussian assumption on the process, the estimation of FPC scores 
can be obtained based on conditional expectation proposed by Mardia et al. (1979)
or Yao et al. (2005). 

Once obtaining $\hat\mu(t)$, $\hat\phi_l(t)$, $\hat\xi_{il}$ and the number of nonzero
eigenvalues $L$, the trajectory of the $i$th patient can be estimated by 
\begin{align}
\label{Equation-estimated trajectory}
\hat X_i(t)=\hat\mu(t)+\sum_{l=1}^L\hat\xi_{il}\hat\phi_l(t).
\end{align}

After selecting the number of nonzero eigenvalues $L$ and estimating
the mean curve $\hat\mu(t)$, the corresponding FPC $\hat\phi_l(t)$ and 
the FPC score $\hat\xi_{il}$ from all available data, we can estimate
missing values and predict future values of individual $i$ using 
the recovered trajectory (\ref{Equation-estimated trajectory})
\begin{align*}
\hat X_i(t)=\hat\mu(t)+\sum_{l=1}^L\hat\xi_{il}\hat\phi_l(t)
\end{align*}
where time $t$ can be any past or future time points.

To evaluate the accuracy in forecasting lung function values, 
the cross validated mean square error can be computed. 
For patients who have more than two observations ($n=60$),
we predict the last visit value by using all previous outcomes of this patient 
and by borrowing the information of all the other patients ($n=117$) using the above FPCA method.
Thus to predict the last value for individual $i$, the training set on which the model is based,
includes the information of all the patients except for the value of the last visit of patient $i$. 
The forecast is given by 
\begin{align*}
\hat X_i^{(-last)}(t)=\hat\mu^{(-last)}(t)+\sum_{l=1}^{L^{(-last)}}\hat\xi^{(-last)}_{il}
\hat\phi^{(-last)}_l(t)
\end{align*}
where $\hat\mu^{(-last)}(t)$ is a local linear estimator of $\mu(t)$ with the last value
of individual $i$ being excluded, $\hat\phi^{(-last)}_l(t)$ is the estimator of $\phi_l(t)$ 
with the last value of individual $i$ being excluded, $\hat\xi^{(-last)}_{il}$
is the estimator of $\phi_l(t)$ with the last value of individual $i$ being excluded,
$L^{(-last)}$ is the number of nonzero eigenvalues selected by CV with the last value 
of individual $i$ being excluded.
Therefore, the estimation of the last value of individual $i$ by using its
own historical information and all the information of others is given by 
\begin{align*}
\hat X_i^{(-last)}(t_{i, last})
=\hat\mu^{(-last)}(t_{i, last})+\sum_{l=1}^{L^{(-last)}}\hat\xi^{(-last)}_{il}\hat\phi^{(-last)}_l(t_{i, last})
\end{align*}
where $t_{i, last}$ is the last visiting time for individual $i$.



To assess the accuracy of the forcasts the mean squared errors  are computed, i.e. 
$$MSE
=\frac1n\sum_{i=1}^n(x_i-\hat x_i)^2.$$
for observations are $x_1,..., x_n$ and corresponding prediction values are 
$\hat x_1,..., \hat x_n$. 

Finally the coefficient of determination $R^2$ which assesses the importance of
the improvement in fit from the simpler model is computed:
$$
R^2=1-\frac{MSE_1}{MSE_0},
$$
where $MSE_1$ and $MSE_0$ are the MSE for the model based on the historical values and 
 the null model, respectively.

Finally we assess whether the biomarkers have an effect on the forcast. We regress the residuals of the predicted values of the three outcomes (FVC, TLC, DLCO), 
i.e. $r_{i,t_{i,last}}=X_i(t_{i,last})-\hat X_i^{(-last)}(t_{i.last})$,
on the observed values of the biomarker $z_{i,t_{i,last}}$
at last time points for the patients who have at least three observations ($n=60$).
Specifically, we consider the simple linear regression
$$r_{i,t_{i,last}}=\alpha+\beta z_{i,t_{i,last}}+\epsilon_i$$
and denote the estimated intercept and slope as $\hat\alpha$ and $\hat\beta$ respectively.
Then the residuals are estimated as 
$\hat r_{i,t_{i, last}}=\hat\alpha+\hat\beta z_{i,t_{i,last}}$.
We then compute our updated predicted value as 
$$\hat X_i(t_{i.last} )= \hat X_i^{(-last)}(t_{i.last})+\hat r_{i,t_{i, last}}$$
Notice that for NT, the logarithm transformation is used.

\section{Results}
\subsection{Linear mixed effects models}

For each combination of lung function index (FVC, TLC, DLCO) and biomarker (TIMP, P3NP, HA, NT), linear mixed model (\ref{Equation-Mixed}) is fitted.
Table \ref{Table-Mixed} shows the $p$-values of the chi-square tests which compare the model with the biomarker to the null model without a biomarker.
It appears that there is not much evidence for an association between the biomarkers and the lung function indexes. Only the effect of P3NP on DLCO is statistically significant at the 5\% level.
\begin{table}[ptbh]
\centering  {
\begin{tabular}
[c]{|c|c|c|c|c|c|}\hline
& TIMP & P3NP & HA & NT  \\\hline
FVC    & 0.25 & 0.21 & 0.68 & 0.99 \\
TLC    & 0.87 & 0.61 & 0.38 & 0.37 \\
DLCO   & 0.13 & 0.02 & 0.85 & 0.52 \\
\hline
\end{tabular}
}
\caption{The $p$-values for the effects of the biomarkers on the outcomes.}
\label{Table-Mixed}
\end{table}

For each lung function biomarker combination, the leave one out cross validated average MSE of the prediction based on the mixed effects model is given in Table \ref{Table-Mixed-MSE}.
Here, also the leave one out cross validated average MSE of 
the prediction is given based on a mixed effects model without the biomarker (null model).
It appears that the mixed effects models including one of the biomarkers only slightly improve the prediction compared to the null model.
The largest improvement is 5.3 \% for the combination of P3NP on DLCO.
\begin{table}[ptbh]
\centering  {
\begin{tabular}{|c|c|c|c|c||c|}\hline
       & TIMP & P3NP & HA  & NT  & null\\\hline
FVC    & 644  & 646  & 654 & 650 & 649\\
TLC    & 405  & 400  & 405 & 401 & 401\\
DLCO   & 306  & 300  & 320 & 318 & 317\\
\hline
\end{tabular}
}
\caption{The cross validated MSE's for predictions based on various model. Null is the model without a biomarker, i.e. only time is included in the mixed effects model}
\label{Table-Mixed-MSE}
\end{table}

\subsection{Functional principal component analysis}
\begin{figure}
  \includegraphics[angle=-90, width=0.88\textwidth]{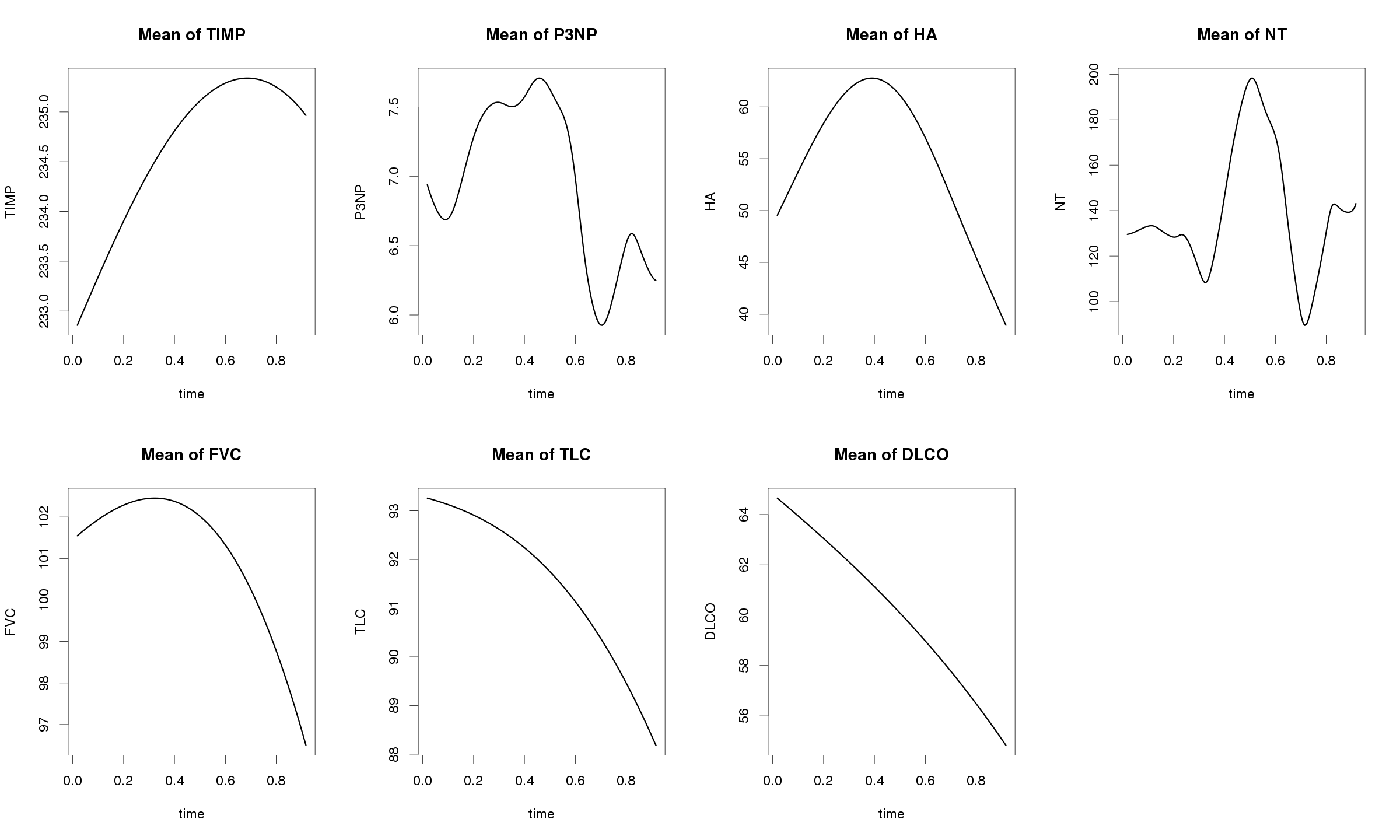}
  \caption{Mean.}
  \label{FigureMean}
\end{figure}

The estimated mean functions of the outcomes (FVC, TLC, DLCO) decrease during the whole period of five years (see Figure \ref{FigureMean}). Note that we  rescaled the 60 months into a [0, 1] interval. All the biomarkers increase in the first two to three years. During the last two to three years, the biomarkers P3NP, HA and NT have a decreasing trend.  At the end of the interval the estimate of the curves for the biomarkers appears to be uncertainty, which is probably caused by the lack of sampling points at the end of the interval  
(see Figure \ref{FigureCleanData}).

The mean function for mRSS has a decreasing trend. Note, however, that the measurement error is around 5 points, 
hence the slight decrease is not medical relevant and we may assume that the mean function is constant over time.

\begin{figure}
  \includegraphics[angle=-90, width=0.88\textwidth]{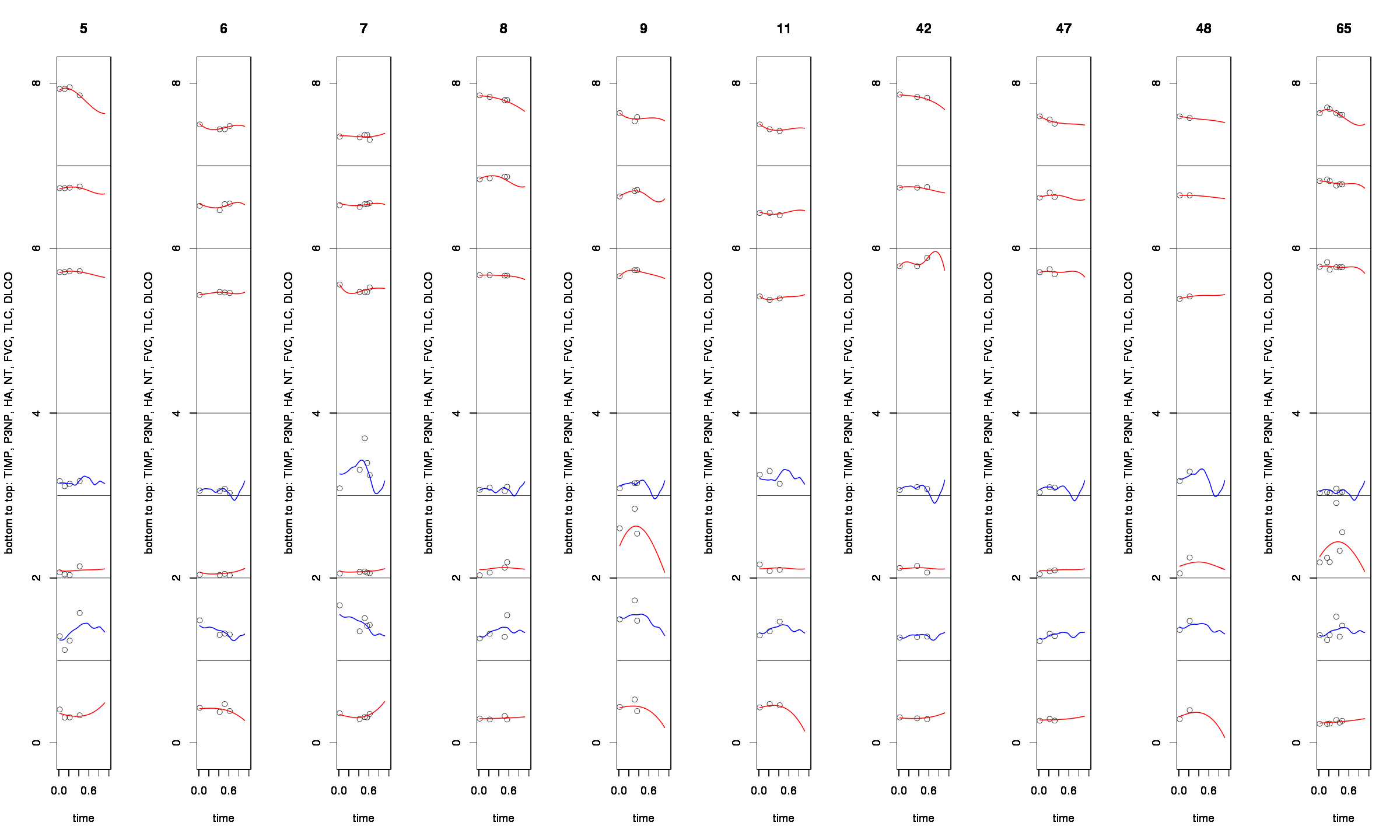}
  \caption{Trajectory.}
  \label{FigureTrajectory1}
\end{figure}

We used two or three principal components to approximate the individual curves.
Figure \ref{FigureTrajectory1} displays the curves  for 
the first 10 patients for TIMP, P3NP, HA, NT, FVC, TLC, and DLCO (from bottom to top).
Overall, the trajectories for each patient captures the behavior
of the observations.
For the batches of blue lines, the recovered cures fluctuate due to
the shape of the mean functions (see Figure \ref{FigureMean}). 

\begin{figure}
  \includegraphics[angle=-90, width=0.88\textwidth]{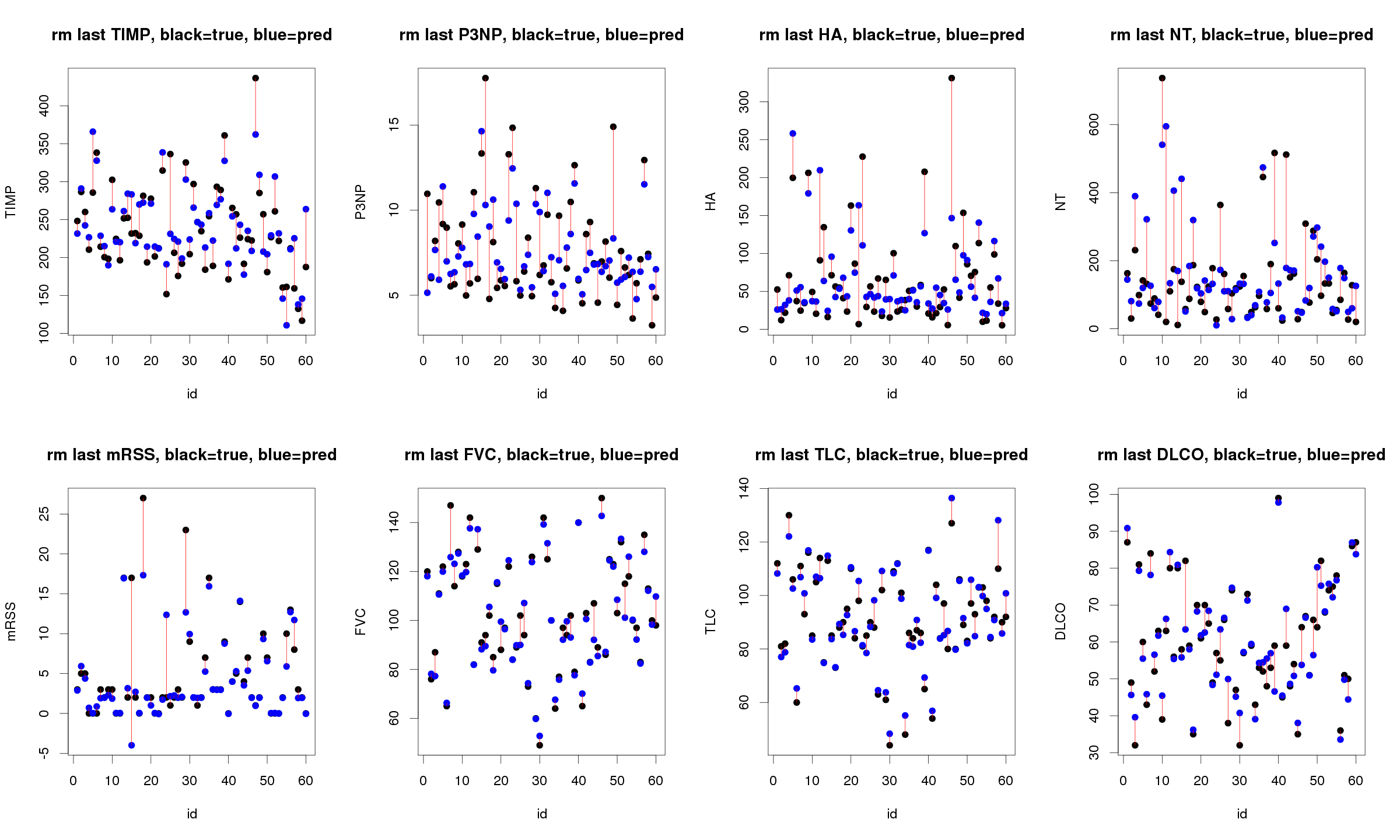}
  \caption{Prediction.}
  \label{FigurePrediction}
\end{figure}

In order to quantify the prediction performance of our FPCA model, 
the predicted values of the last visit are computed.
We only predicted the last values of the patients who have at least three measurements ($n=60$ patients).
Figure \ref{FigurePrediction} shows the prediction results.
Here the blue points are the predicted values, and
the black points are the observed values. The length of the red bars are the prediction errors.
We observe that in general the predictions are good, however for patients who had a long period 
between the last and the previous visit and for time points in the second half of the time interval, 
the prediction error appears to be larger. 
This is probably due to lack of information to predict this value.

The cross validated MSE's for the prediction are shown in Table \ref{TableMSE}.
The cross validated MSE for the null model, i.e.\ prediction
of the last value by the average of the 60 patients is also provided.
Compared to the null model, our model has a much smaller cross validated MSE. In addition the improvement in prediction using historical data is better (90\% reduction in MSE) than the improvement in prediction by using the biomarkers in the mixed-effects model (1\% reduction in MSE) (Table \ref{Table-Mixed-MSE}).
Table \ref{TableMSE} also shows the MSE for the subgroup of patients who have a time interval of less than six months between the last and previous visit (near) (41 out of 60 patients) and the remaining patients (far) (19 out of 60 patients). We also considered the cross validated MSE in the subgroup of patients who have their last visit in the first 
24 months (early) (36 out of 60 patients) and the remaining patients (late) (24 out of 60 patients).
The MSE for ``near'' and ``early'' groups are smaller for most outcomes than for ``far'' and ``late'' group, because for these groups more information (more visits and more patients) is available.
For TLC, there is not much difference between ``early'' and ``late'' groups. However, there is a difference between ``near'' and ``far'' groups, where the "far" group performs better. The reason is that for the ``near'' group the variation between the last two observations 
in each individual is quite large. Our model has difficulties in capturing this since we used the B-spline basis.
Finally the coefficient of determinations are given 
in Table \ref{TableMSE}. Here the smallest coefficient is 0.87 (DLCO).

\begin{table}[ptbh]
\centering  {\small
\begin{tabular}
[c]{|c|c|c|cc|cc|c|}\hline
& \multicolumn{6}{|c|}{MSE} & R$^2$\\\hline
           & null & full & near & far & early & late &   full \\
\# patients & 60   & 60  & 41    & 19      & 36        & 25       & 60\\\hline
FVC  & 538   & 39  & 32       & 53      & 23      & 62  & 0.93 \\
TLC  & 327   & 27  & 31       & 20      & 28      & 27  & 0.92 \\
DLCO & 244   & 32  & 22       & 55      & 21      & 49  & 0.87 \\\hline
\end{tabular}
}\caption{Results of forecasting the outcomes of 60 patients using functional data analysis. 
Reported are MSE and R$^2$. 
The ``null'' model is the prediction of the last value using the average of the 60 patients.
``near'' corresponds to the group which have their last observation within a period of six months after the previous visit, ``far'' comprises the rest of the patients.
``early''  corresponds to the subgroup of patients which has the last observation time within the first 
24 months of disease onset, ``late'' comprises the rest of the patients.
}%
\label{TableMSE}
\end{table}

Finally we investigated whether the biomarkers can further improve the prediction performance of our model. 
A linear regression model was fitted where the residuals of our model are regressed on the
current biomarker values.
The $p$-values for the regression coefficients are shown in Table \ref{TablePvalues}.
It appears that the biomarkers, P3NP, HA and NT are associated with the prediction of DLCO based on the historical data. 
The biomarker NT is associated with the prediction of FVC and TLC based on their historical values.
The MSEs and the R$^2$ values for the updated predictions are also given in Table \ref{TablePvalues}.
It appears that the predictions are only slightly improved. 
Thus the biomarkers do not improve the prediction either when added to the null model or to the model based
on the historical data.
\begin{table}[ptbh]
\centering  {\small
\begin{tabular}
[c]{|c|c|c|c|c|}\hline
& TIMP & P3NP & HA & NT\\\hline
 \multicolumn{5}{|c|}{p-values}\\\hline
FVC & 0.807 & 0.365 & 0.578 & 0.034 \\
TLC & 0.589 & 0.865 & 0.627 & 0.032 \\
DLCO & 0.819 & 0.013 & 0.012 & 0.069 \\\hline
 \multicolumn{5}{|c|}{MSE}\\\hline
FVC (39) & 38 & 38 & 38 & 35 \\
TLC (27) & 27 & 27 & 27 & 25 \\
DLCO (32)& 32 & 29 & 29 & 31 \\\hline
 \multicolumn{5}{|c|}{R$^2$}\\\hline
FVC  & 0.00 & 0.04 & 0.03 & 0.10 \\
TLC  & 0.00 & 0.00 & 0.00 & 0.07 \\
DLCO & 0.00 & 0.09 & 0.09 & 0.03 \\\hline
 \multicolumn{5}{|c|}{cross validated MSE }\\\hline
FVC (39) & 40 & 40 & 40 & 39 \\
TLC (27) & 29 & 29 & 30 & 28 \\
DLCO (32)& 35 & 32 & 31 & 34 \\\hline
\end{tabular}
}\caption{
Results of the methods which include the biomarker information of the last time point in the forecast model based on patients historical observations. Reported are p-values to assess association and MSE and R$^2$ to assess prediction performance. }

 \label{TablePvalues}
\end{table}


\section{Discussion}
In this paper we considered two methodological approaches to predict new lung function values using candidate biomarkers and patient's historical information on these lung functions measured for Scleroderma patients. We considered linear mixed models to assess the predictive value of the biomarkers. Linear mixed models are popular for the analysis of data from epidemiological studies. However data from EHR are irregular. Indeed in our study we could only analyze data from 73 patients with regular observations, while we had 117 patients in total. Unfortunately to forecast new values based on historical information using mixed models was not possible since we do not have sufficient patients with more than two regular observed values. Therefore for forecasting based on historical data, we considered functional data analysis methods.

With regards to the biomarkers we obtained some evidence for association between the biomarkers P3NP, HA and NT with the three lung functions DLCO (all three) and FVC and TLC (HA and NT). However the added predictive value to the null model and to the model with historical data was small. 
It might be that our models are too simple to capture the relationship between the biomarkers and the outcomes.
More advanced methods are needed to model non linear relationships or a delayed effect of the biomarker on the outcome. Unfortunately, the dataset was too small to investigate this kind of models. 

Forecasting based on historical data using functional data analysis methods performed well. The cross validated MSEs of these models were much smaller than the cross validated MSEs of the null model. Especially for time intervals less than or equal to six months and in the first two years after disease onset, estimation of the lung function values was good. 
It is expected that when more data are available even better predictions can be obtained for the whole disease trajectory if the time gaps are not too large. 
Further improvements of prediction accuracy might be obtained by including the biomarker trajectories in the regression part instead of only the last values, by joint modelling of the three correlated lung functions and by considering a model based on a mixture approach with as unobserved status the membership of the group which will develop the disease (30\% of the patients).  

A drawback of functional data analysis methods is that it is not likelihood based. Hence parameter estimates might be biased if we have data missing at random. In this study we have no drop out so we expect that our estimates are unbiased. For studies with drop outs, inverse probability weighting 
(Kurland and Heagerty, 2006) might be a solution.

To conclude we successfully applied functional data analyses methods to forecast new lung function values based on historical data. These methods are relevant since lung function measurements are expensive. Moreover applications of these methods might result in less visits to the hospital which improves quality of life of Scleroderma patients. 
Further in the future improvements might be obtained by development of new methodology for a larger group of patients with a longer follow up time.

\end{document}